\documentclass[12pt,tightenlines,eqsecnum,floats,aps,amsmath,amssymb,nofootinbib,prd,showpacs]{revtex4}
\usepackage{setspace}
\usepackage{amsmath,amssymb,amsfonts,amsthm}
\usepackage{graphicx}
\usepackage{enumerate} 
\usepackage{colordvi} 

\def\be{\begin{equation}}
\def\ee{\end{equation}}
\def\ba{\begin{eqnarray}}
\def\ea{\end{eqnarray}}

\def\cm{\rm cm}

\def\pphi{p_{(\phi)}}
\def\lp{{\ell}_{\rm Pl}}
\def\S{\mathbb{S}}
\def\L{\mathrm{L}}
\def\B{\mathrm{B}}
\def\rcr{\rho_{\mathrm{crit}}}
\def\rmin{\rho_{\mathrm{min}}}
\def\rmax{\rho_{\mathrm{max}}}
\def\b{$\bullet\,\,\,\, $}
\def\f{\frac}
\def\dd{\textrm{d}}

\def\WDW{WDW\,\,}

\usepackage{enumerate}

\usepackage{colordvi}
\newcounter{mnotecount}[section]

\newcommand{\comment}[1]{}

\begin{document}

\title{The Big Bang and the Quantum}

\author{Abhay Ashtekar}
\affiliation{Institute for Gravitation and the Cosmos \& Physics
Department\\ Penn State, University Park, PA 16802, U.S.A.}

\begin{abstract}

This short review is addressed to cosmologists.%
\footnote{Plenary talk at the \emph{Invisible Universe} conference
held in Paris in July 2009. To appear in the Proceedings, edited by
J. M. Alimi et al  (AIP Publications)}

General relativity predicts that space-time comes to an end and
physics comes to a halt at the big-bang. Recent developments in loop
quantum cosmology have shown that these predictions cannot be
trusted. Quantum geometry effects can resolve singularities, thereby
opening new vistas. Examples are: The big bang is replaced by a
quantum bounce; the `horizon problem' disappears; immediately after
the big bounce, there is a super-inflationary phase with its own
phenomenological ramifications; and,
in presence of a standard inflaton potential, initial conditions are
naturally set for a long, slow roll inflation independently of what
happens in the pre-big bang branch. 

\end{abstract}


\maketitle

\section{Introduction}
 \label{s1}

At this conference we heard of the spectacular progress that has
occurred in observational cosmology in recent years. We also learned
about the upcoming missions that are poised to provide new data to
further constrain or even rule out leading theoretical models. These
advances have been and continue to be the engines that drive
contemporary cosmology. They have brought to forefront the
astonishing success of the Friedmann, Lema\^{\i}tre, Robertson,
Walker (FLRW) models, and perturbations thereof. Indeed, it appears
that the rich data that we now have, and are likely to accumulate in
the near future, would be adequately described by these simple
applications of general relativity and quantum field theory on the
resulting cosmological backgrounds.

However these theories are conceptually incomplete. They assume that
the universe begins with a big bang at which matter densities and
space-time curvature become infinite. With inflationary scenarios
there were initial hopes that perhaps the big-bang singularity could
be avoided because the inflaton fails to satisfy the strong energy
condition often used in the singularity theorems of general
relativity. However, Borde, Guth and Vilenkin \cite{bgv} soon
established that this hope was misplaced. Inflation can be eternal
in the future but, if we go back in time using Einstein's equations,
one again finds that the space-time ends and physics simply comes to
a halt at the big bang. But all our experience with fundamental
physics suggests that this cannot be the situation in the real
world. This must be a prediction of a theory that has been pushed
well beyond the domain of its validity. To know what \emph{really}
happened near the putative big bang, we must work with a genuine
unification of general relativity and quantum physics, an
unification which does not pre-suppose that space-time is a smooth
continuum, and which can encompass the rich non-linear structure of
strong gravity that lies beyond the scope of perturbation theory.

But the burden on this desired unification is heavy. In the context
of cosmology there is a long list of fundamental questions that must
be satisfactorily addressed by such a theory. Here is an
illustrative list encompassing some of the contemporary issues.
\medskip
\begin{quote}
\noindent$\bullet$ If general relativity is transcended, how close
to the putative big bang does a smooth space-time of Einstein's make
sense? In particular, can one show from first principles that this
approximation is valid at the onset of inflation?\\
$\bullet$ Is the big-bang singularity naturally resolved by the
quantum version of Einstein's equations? Or, is some external input
such as a new principle or a boundary condition at the big bang
essential? An outstanding example of such an external input is the
Hartle-Hawking proposal \cite{hh}.\\
$\bullet$ Is the quantum evolution across the `singularity'
deterministic? One needs a fully non-perturbative framework to
answer this question in the affirmative. In the pre-big-bang
\cite{pbb} and ekpyrotic/cyclic \cite{ekp1,ekp2} scenarios, for
example, so far the answer is in the negative because these theories
pre-suppose the space-time continuum of general relativity and
this approximation fails at the big-bang.\\
$\bullet$ If the singularity is resolved, what is on the `other
side'? Is there just a `quantum foam', far removed from any
classical space-time, or, is there another large, classical
universe?
\end{quote}

Such questions are fundamental and must be faced squarely because
the resolution of classical singularities can profoundly shift the
paradigm underlying contemporary cosmology. This in itself makes it
imperative that we understand the quantum nature of the big bang.
But there could also be another rich pay-off: the new paradigm may
enable us to address open issues that are observationally
significant. For example, if there is a classical pre-big-bang
branch to the universe, the horizon problem would disappear and the
observed large scale homogeneity could be simply a consequence of
the fact that even the most distant parts of the universe would have
been in causal contact in the past. If there is a pre-big-bang
branch, we would not have to specify the initial conditions for
perturbations on the singularity, where the applicability of current
theories is least reliable. There would be more natural ways of
specifying these conditions which, in turn, may well lead to small,
potentially observable deviations from current predictions. Finally,
our experience with general relativity itself suggests that, once
the physics of the Planck regime is well understood, we may be
handed with novel predictions of central importance to the next
generation of astrophysicists and cosmologists.

Loop quantum gravity (LQG) \cite{alrev,crbook,ttbook} is well suited
to embark on this mission because it does not pre-suppose a
classical space-time --it is background independent. At a
fundamental level, everything, \emph{including geometry,} is
described in the paradigm of quantum physics
\cite{almmt,rs,al5,alvol}. Classical space-times emerge only on
coarse graining of semi-classical quantum states. Finally, since the
approach is fully non-perturbative, it is well suited for the strong
field regime near the putative big bang.

In this chapter I will discuss loop quantum cosmology (LQC),
the application of the principles of LQG to cosmology
\cite{mb-rev,aa-badhonef}. Initial ideas appeared in
\cite{mb1,abl} and were developed in detail for a variety of
cosmological models in
\cite{aps1,aps2,aps3,ps,apsv,kv,acs,cs1,bp,ap,aps4,hybrid,awe2,awe3,cs2}.
\emph{In all cases, the big-bang and big-crunch singularities
are resolved in a direct physical sense.} The resulting Planck
scale physics has been explored using analytical and numerical
solutions to the quantum Einstein equations as well as
effective equations which capture the leading quantum
corrections. Singularity theorems are avoided \emph{not}
because one uses matter violating energy conditions. Indeed, in
the models that have been studied in most detail, \emph{all
energy conditions hold}. The theorems are inapplicable because
quantum geometry effects modify Einstein equations
themselves.

Physically, quantum geometry gives rise to a new repulsive
force. This force is \emph{utterly} negligible under normal
circumstances. It is only when the matter density becomes about
1\% of the Planck density or curvature approaches $1/\lp^2$
that the repulsive force ---and hence the deviation from
classical general relativity--- becomes significant. But then
the repulsive force rises \emph{very} quickly, overwhelms
classical attraction and causes a quantum bounce. The density
and curvature start falling and once they are below the scales
just mentioned, the force again becomes negligible and
classical general relativity again becomes an excellent
approximation.

Immediately to the future of the bounce there is a robust phase
of super-inflation which is not encountered in general
relativity \cite{mb2,ps2}. But it is short lived and in absence
of a suitable inflaton potential it does not yield a sufficient
number of e-foldings. However, in presence of suitable
potentials ---such as $m^2\phi^2$--- super-inflation
\emph{funnels the phase space trajectories to initial
conditions which virtually guarantee a slow roll inflation with
60 or more e-foldings} \cite{as}. This is in striking contrast
to what happens in general relativity where it has been argued
\cite{gt} that the probability of $N$ e-folding decreases as
$e^{-3N}$. The super-inflationary phase is also likely to have
other phenomenological consequences ---such as production of
gravitational waves--- that are being analyzed.

The article is organized as follows. Section \ref{s2} lays out the
conceptual setting and section \ref{s3} provides a bird's eye view
of LQC through illustrative results. We will see that not only has
LQC answered many of the long standing questions but it has also
opened new vistas. Section \ref{s4} summarizes the origin of the
novel predictions and places them in a broader perspective.

\section{Conceptual Setting}
\label{s2}
\begin{figure}[]
  \begin{center}
$a)$\hspace{8cm}$b)$
    \includegraphics[width=3.2in,angle=0]{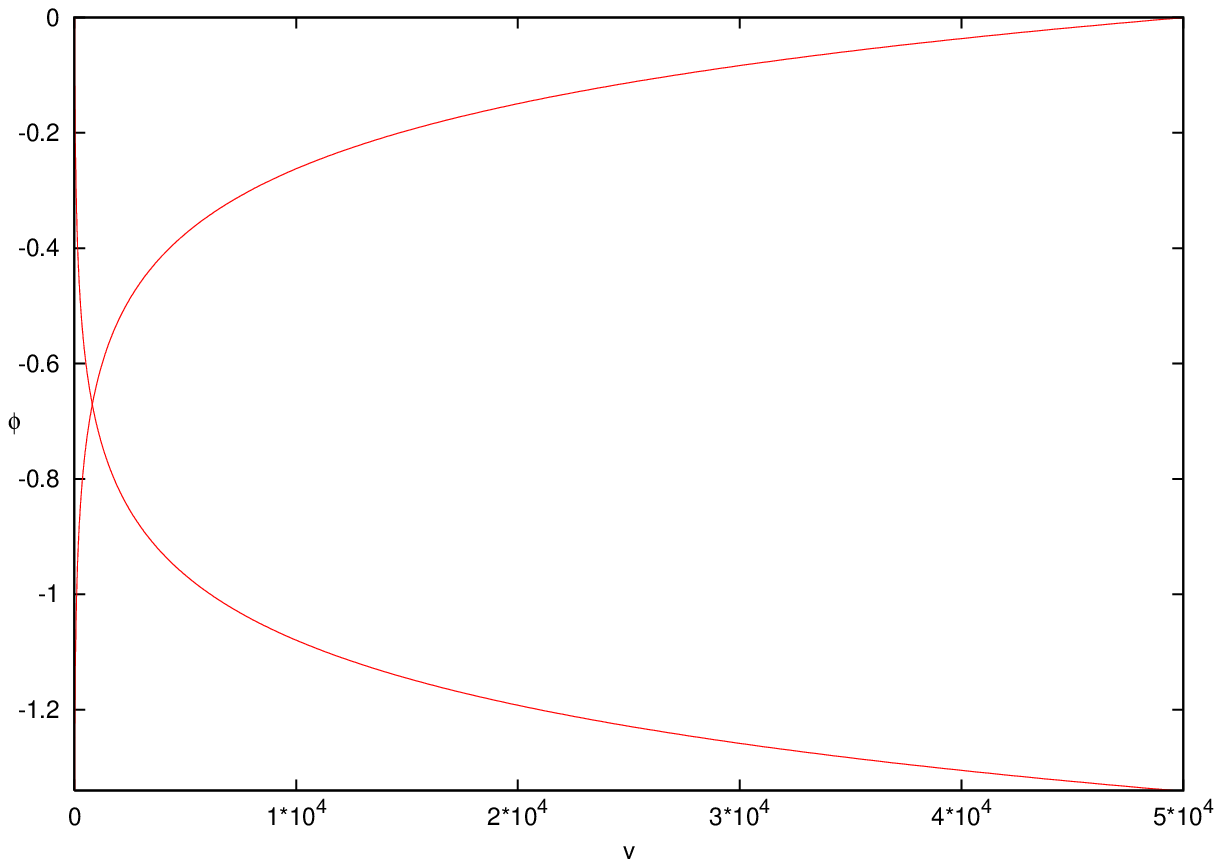}
\includegraphics[width=3.2in,angle=0]{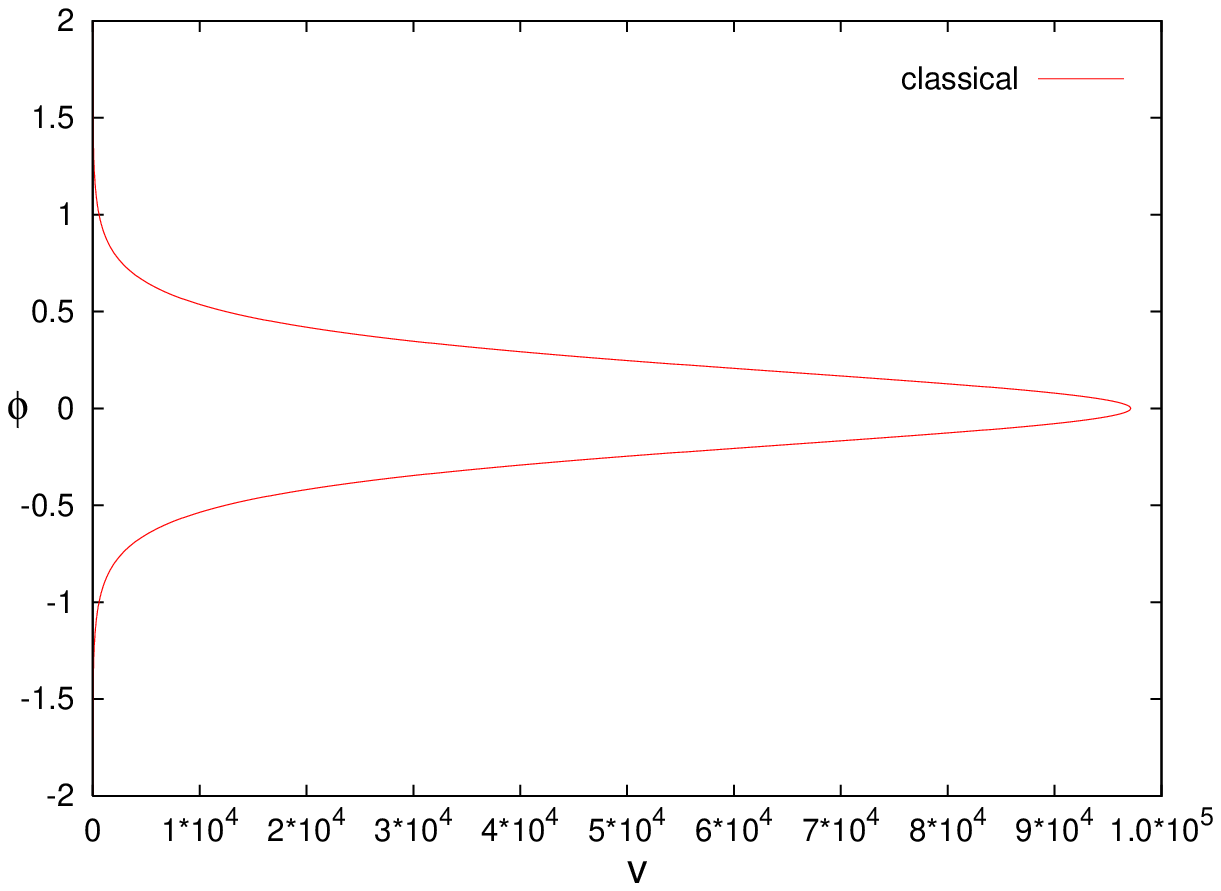}
\caption{Dynamics of FLRW universes with zero cosmological
constant and a massless scalar field. Classical trajectories are
plotted in the $v-\phi$ plane, where $v \sim a^3$ denotes the volume
and $\phi$ the scalar field. Following the convention in general
relativity, the (emergent) time variable $\phi$ is plotted along the y-axis.
$a)$ k=0 trajectories. $b)$ A k=1 trajectory. In the k=0 case, $v$
is the physical volume of a fiducial cell/box; $v\sim a^3$ where $a$
is the scale factor. The final physical results are of course
insensitive to the choice of this cell.}\label{class}
\end{center}
\end{figure}

To set the stage, let me use the simplest cosmological models: FLRW
space-times with a massless scalar field. These models are
instructive because in classical general relativity all their
solutions have a big-bang (and/or big-crunch) singularity.
Therefore, a quantum resolution of these singularities is
non-trivial. Furthermore, it is not difficult to incorporate
potentials, additional matter fields and anisotropies.

Figure \ref{class} illustrates classical dynamics for k=0 and
k=1 models without a cosmological constant. $\phi$ is the
massless scalar field while $v\sim a^3$ denotes the
\emph{physical} volume of the universe in the k=1 case and of a
fixed fiducial cell in the k=0 case. If k=0 there are two
classes of trajectories. In one the universe begins with a
big-bang and expands continuously, while in the other, it
contracts continuously into a big crunch. In the k=1 case, the
universe begins with a big bang, expands to a maximum volume
and then undergoes a recollapse to a big crunch singularity.
Now, in quantum gravity, one does not have a single space-time
in the background but rather a probability amplitude for
various space-time geometries. Therefore, unlike in the
classical theory, one cannot readily use, e.g.,  the proper
time along a family of preferred observers as a clock. However,
along each dynamical trajectory, the massless scalar field
$\phi$ is monotonic. Therefore, it serves as a good clock or
`emergent time' with respect to which the physical degrees of
freedom ---the matter density or the volume, anisotropies,
curvature scalars and other matter fields, if any--- evolve.
Note incidentally that, in the k=1 case, because of the
classical recollapse, volume (or, the scale factor) is
double-valued along any dynamical trajectory. Therefore it
cannot serve as a global clock variable, while the scalar field
does fulfill this role.

If one has a full quantum theory, one can proceed as follows. Choose
a classical solution, i.e. a dynamical trajectory in the $v,\,\phi$
plane. The momentum $\pphi$ conjugate to $\phi$ is a constant of
motion. Let us suppose its value on our trajectory is $\pphi =
p_{(\phi)}^\star$. Next, choose a point $(v^\star, \phi^\star)$ on
the trajectory where the matter density and space-time curvature are
low. This point describes the state of the FLRW universe at a late
time when general relativity is expected to be valid. At the `time'
$\phi=\phi^\star$ construct a wave packet which is sharply peaked at
$v=v^\star$ and $\pphi = p_{(\phi)}^\star$ and evolve it backward
and forward in (the scalar field ) time. We are then led to two
questions:
\smallskip

i) \emph{The infrared issue:} Does the wave packet remain peaked on
the classical trajectory in the low curvature regime? Or, do quantum
geometry effects accumulate over the cosmological time scales,
causing noticeable deviations from classical general relativity? In
particular, in the k=1 case, is there a recollapse and if so for
large universes does the value $V_{\rm max}$ of maximum volume agree
with that predicted by general relativity \cite{gu}?

ii) \emph{The ultraviolet issue:} What is the behavior of the
quantum state when the curvature grows and enters the Planck regime?
Is the big-bang singularity resolved without any extra input? Or, do
we need to supplement dynamics with a new principle, such as the
Hartle-Hawking `no boundary proposal' \cite{hh}? What about the
big-crunch?  \smallskip

By their very construction, perturbative and effective descriptions
have no problem with the first requirement. However, physically
their implications can not be trusted at the Planck scale and
mathematically they generally fail to provide a deterministic
evolution across the putative singularity. Since the
non-perturbative approaches often start from deeper ideas, it is
conceivable that they could lead to new structures at the Planck
scale which modify the classical dynamics and resolve the big-bang
singularity. But once unleashed, do these new quantum effects
naturally `turn-off' sufficiently fast, away from the Planck regime?
The universe has had some \emph{14 billion years} to evolve since
the putative big bang and even minutest quantum corrections could
accumulate over this huge time period leading to observable
departures from dynamics predicted by general relativity. Thus, the
challenge to quantum gravity theories is to first create huge
quantum effects that are capable of overwhelming the extreme
gravitational attraction produced by matter densities of some
$10^{94}\, {\rm gms/cc}$ near the big bang, and then switching them
off with extreme rapidity as the matter density falls below this
Planck scale. This is a huge burden!

The question then is: How do various approaches fare with respect to
these questions? The older quantum cosmology ---the Wheeler-DeWitt
(WDW) theory--- passes the infra-red test with flying colors. But
unfortunately the state follows the classical trajectory into the
big bang (and in the k=1 case also the big crunch) singularity. The
singularity is not resolved because \emph{expectation values of
density and curvature continue to diverge} in epochs when their
classical counterparts do \cite{aps2,aps3}.

For a number of years, the failure of the \WDW theory to naturally
resolve the big bang singularity was taken to mean that quantum
cosmology cannot, by itself, shed significant light on the quantum
nature of the big bang. Indeed, for systems with a finite number
of degrees of freedom we have the von Neumann uniqueness theorem
which guarantees that quantum kinematics is unique. The only
freedom we have is in factor ordering and this was deemed
insufficient to alter the status-quo provided by the \WDW theory.

The situation changed dramatically in LQG. In contrast to the \WDW
theory, a well established, rigorous kinematical framework \emph{is}
available in full LQG \cite{almmt,alrev,crbook,ttbook}. Furthermore,
this framework is uniquely singled out by the requirement of
diffeomorphism invariance (or background independence)
\cite{lost,cf}. If one mimics it in symmetry reduced models, one
finds that a key assumption of the von-Neumann theorem is violated.
As a result, \emph{one is led to new quantum mechanics} \cite{abl}!
This quantum theory is inequivalent to the \WDW theory already at
the kinematic level. Quantum dynamics built in this new arena agrees
with the \WDW theory in `tame' situations but differs dramatically
in the Planck regime, leading to a natural resolution of the big
bang and the big crunch singularities.

\section{Loop Quantum Cosmology}
\label{s3}

This section is divided into three parts. In the first I consider
the FLRW models without a cosmological constant in some detail, in
the second I summarize the more general situation, and in the third
I discuss the phenomenon of super inflation and its surprising
implications for inflationary scenarios.

\subsection{Singularity resolution in the FLRW Models}
\label{s3.1}

The main LQC results can be summarized as follows
\cite{aps1,aps2,aps3,kv,acs,cs1,apsv,bp,ap,aps4}.\smallskip

\begin{figure}[]
  \begin{center}
    $a)$\hspace{8cm}$b)$
    \includegraphics[width=3.2in,angle=0]{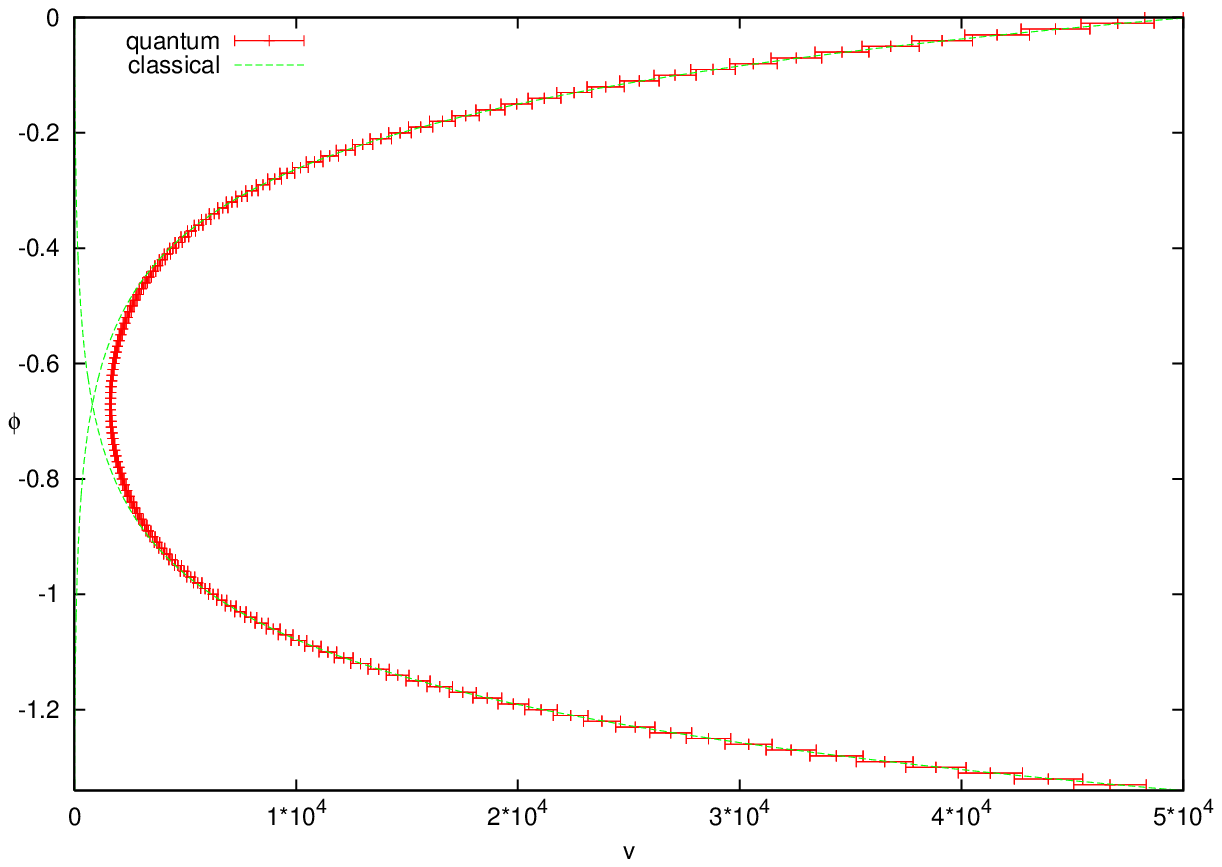}
    \includegraphics[width=3.2in,angle=0]{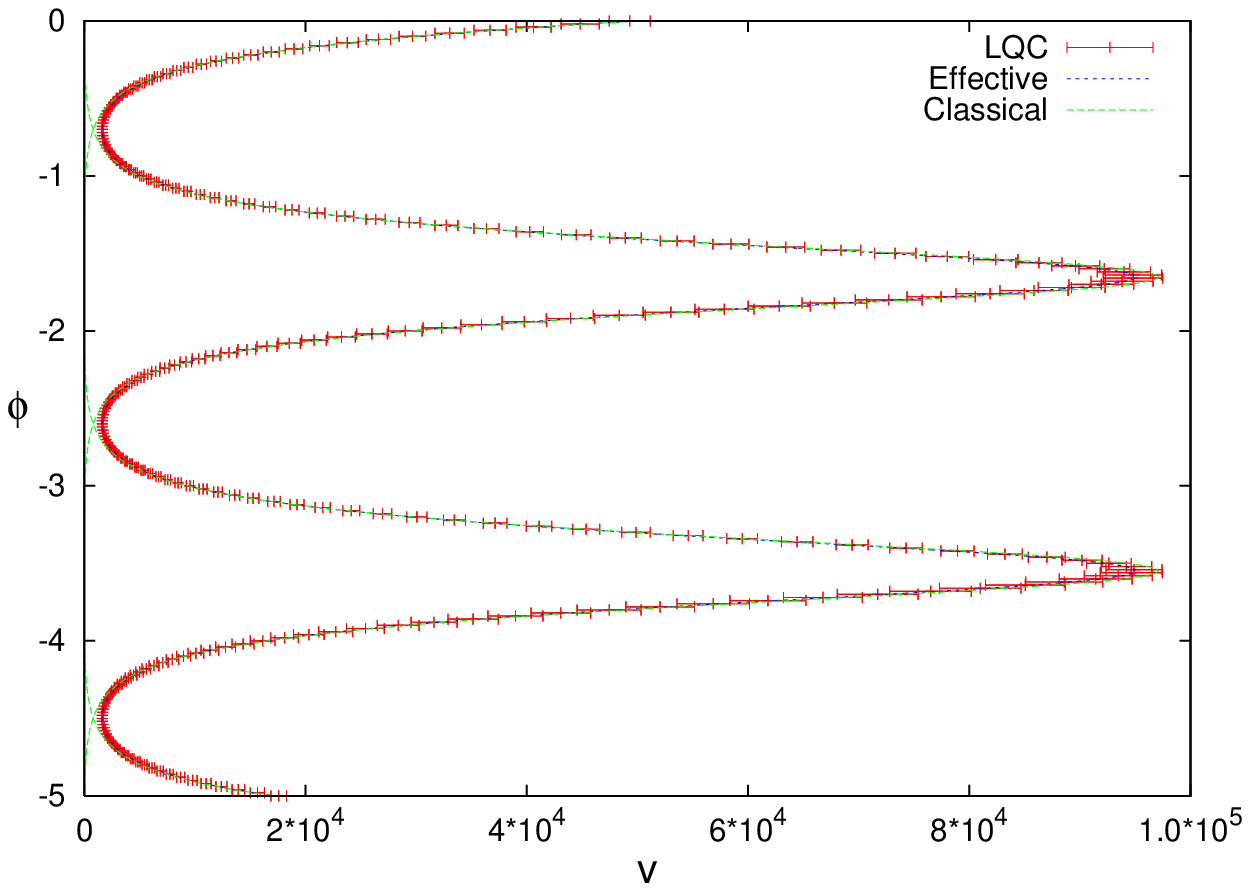}
\caption{In the LQC evolution of models under consideration, the
big bang and big crunch singularities are replaced by quantum
bounces. Expectation values and dispersion of the volume operator
are compared with the classical trajectory and the
trajectory from effective Friedmann dynamics. The classical
trajectory deviates significantly from the quantum evolution at
the Planck scale and evolves into singularities. The effective
trajectory provides an excellent approximation to quantum
evolution at all scales. \,\, $a)$ The k=0 case. In the backward
evolution, i.e., as $\phi$ decreases, the wave function
follows our post big-bang branch at low densities and curvatures
but undergoes a quantum bounce at matter density $\rho \sim 0.41
\rho_{\rm Pl}$ and joins on to the classical trajectory that was
contracting to the future.
$b)$ The k=1 case. The quantum bounce occurs again at $\rho \sim 0.41
\rho_{\rm Pl}$. Since the big bang and the big crunch are replaced
by quantum bounces and the classical re-collapse survives, the
evolution undergoes cycles. In these simulations, $p_{(\phi)}^\star
= 5\times 10^3$, $\Delta p_{(\phi)}/p_{(\phi)}^\star = 0.018$,
and $v^\star = 5\times 10^4$.} \label{fig:lqc1}
\end{center}
\end{figure}

Let us first consider the k=0 model without a cosmological constant.
Following the strategy outlined in section \ref{s2}, let us fix a
point at a late time on the trajectory corresponding to an expanding
classical universe, construct a Gaussian wave function which is
sharply peaked at that point, and evolve it using the LQC
Hamiltonian constraint. One then finds the following
\cite{aps3}.\smallskip

\b The wave packet remains sharply peaked on the classical
trajectory so long as the matter density $\rho$ remains below 1\% of
the Planck density $\rho_{\rm Pl}$. Thus, as in the \WDW theory, the
LQC evolution meets the infra-red challenge successfully.

\b Let us evolve the quantum state back in time, toward the
singularity. In the classical solution scalar curvature and the
matter energy density keep increasing and eventually diverge at the
big bang. The situation is very different in LQC. As mentioned in
section \ref{s1}, once the density and curvature enter the Planck
scale \emph{quantum geometry effects become dominant creating an
effective repulsive force which rises very quickly, overwhelms the
classical gravitational attraction, and causes a bounce thereby
resolving the big bang singularity.} (See Fig \ref{fig:lqc1}.)
Numerical simulations show that the density acquires its maximum
value $\rmax \approx 0.41 \rho_{\mathrm{Pl}}$ at the bounce point.

\b Although in the Planck regime the peak of the wave function
deviates very substantially from the general relativistic trajectory
of figure \ref{class}, it follows an effective trajectory  with very
small fluctuations. This effective trajectory was derived using
techniques from geometric quantum mechanics \cite{jw,vt}. The
effective equations it satisfies incorporate the leading corrections
from quantum geometry which modify the left hand side of Einstein's
equations. However, to facilitate comparison with the standard form
of Einstein's equations, one moves this correction to the right side
through an algebraic manipulation. Then, one finds that the
Friedmann equation
\be \label{fe} \left( \f{\dot{a}}{a} \right)^2 = \f{8\pi G}{3}\,
\rho\, . \ee
of classical general relativity is replaced by
\be \label{lqc-fe}\left( \f{\dot{a}}{a} \right)^2 = \f{8\pi
G}{3}\,\rho\, \big(1 - \f{\rho}{\rcr}\big) \, . \ee
Here, as usual `dot' refers to the derivative with respect to
\emph{proper time} and $\rcr= {\sqrt{3}}/{32\pi^2\gamma^3G^2\hbar}$,
where $\gamma$ is the Barbero-Immirzi parameter of LQG (whose value
$\gamma \sim 0.24$ is determined by the black hole entropy
calculations in LQG). By plugging in numbers one finds $\rcr \approx
0.41 \rho_{\rm Pl}$. Thus, $\rcr \approx \rmax$, found in numerical
simulations. Furthermore, one can show analytically \cite{acs} that
the eigenvalues of the density operator on the \emph{physical}
quantum states are bounded above by $\rho_{\rm sup}$, also given by
$\rho_{\rm sup}= {\sqrt{3}}/{32\pi^2\gamma^3G^2\hbar}$.  Thus,
\emph{there is an excellent match between the quantum theory which
provides} $\rho_{\rm sup}$ \cite{acs}, \emph{the effective equations
which provide} $\rcr$ \cite{aps3,vt} \emph{and numerical simulations
which provide $\rmax$} \cite{aps3}.

\b In classical general relativity the right side,\, $8\pi G
\rho/3$,\, of the Friedmann equation is positive, whence $\dot{a}$
cannot vanish; the universe either expands forever from the big bang
or contracts forever ending in the big crunch. In the LQC effective
equation, on the other hand, $\dot{a}$ vanishes when $\rho=\rcr$ at
which a quantum bounce occurs: To the past of this event, the
universe contracts while to the future, it expands. This is possible
because the LQC correction $\rho/\rcr$ \emph{naturally} comes with a
negative sign. This is non-trivial. In the standard brane world
scenario, for example, Friedmann equation also receives a
$\rho/\rcr$ correction but it comes with a positive sign (unless one
artificially makes the brane tension negative) whence the
singularity is not resolved.

\b Consider the standard inflationary scenario for the $m^2\phi^2$
potential with phenomenologically determined values of $m$. Then the
standard initial conditions at the onset of inflation are such that
the quantum correction $\rho/\rcr$ is of the order $10^{-11}$, and
hence completely negligible. Thus, LQC calculations provide \emph{an
a priori justification} for using classical general relativity
during inflation.\\

In the closed, k=1 model, the situation is similar but there are two
additional noteworthy features \cite{apsv}. Although they are not
important from phenomenological considerations, they reveal
surprising properties of the domain of applicability of classical
general relativity.

\b To start with, classical general relativity is again an excellent
approximation to the LQC evolution till matter density $\rho$
becomes about 1\% of the Planck density $\rho_{\rm Pl}$ but, as the
density increases further, the two evolutions start diverging
rapidly. Again, quantum geometry effects become dominant creating an
effective repulsive force which rises very quickly, overwhelms the
classical gravitational attraction, and causes a bounce thereby
resolving both the big bang and the big crunch singularities.
Surprisingly these considerations apply \emph{even to universes
whose maximum radius $a_{\rm max}$ is only} $23 \lp$. For these
universes, general relativity is a very good approximation in the
range $8 \lp < a < 23\lp$! The matter density acquires its minimum
value $\rmin$ at the recollapse. The classical prediction $\rmin =
3/8\pi Ga^2_{\rm max}$ is correct to one part in $10^{5}$. It is
rather astonishing that general relativity is so accurate even in
situations where one would have expected quantum corrections to
dominate.

\b On the other hand there are also situations in which one
would have at first expected general relativity valid, where
quantum corrections dominate! Recall that the volume of the
closed universe acquires its minimum value $V_{\rm min}$ at the
quantum bounce. $V_{\rm min}$ scales linearly with
$p_{(\phi)}$. Consequently, $V_{\rm min}$ can be \emph{much}
larger than the Planck size. Consider for example a quantum
state describing a universe which attains a maximum radius of a
megaparsec. Then the quantum bounce occurs when the volume
reaches the value $V_{\rm min} \approx 5.7 \times 10^{16}\,
{\cm}^3$, \emph{some $10^{115}$ times the Planck volume.} A
more realistic universe would have a much larger maximum radius
and hence $p_{(\phi)}$ and then $V_{\rm min}$ would be much
lager! Although it is at first surprising that quantum effects
can dominate when the universe is so large, a moment's
reflection shows that deviations from the classical behavior
are triggered when the density or curvature reaches the Planck
scale. The size of the volume by itself is not the relevant
factor for quantum gravity effects.\\

Thus, in the simplest k=0 and k=1 FLRW models, classical
singularities are replaced by quantum bounces and LQC provides
a rather detailed picture of the physics in the Planck regime.
The situation is essentially the same in the open, k=-1 model
\cite{kv}. (Although the detailed analysis in \cite{kv} has
certain drawbacks, they can be overcome using techniques
introduced in \cite{awe3}.) Furthermore, the singularity
resolution does not cause infra-red problems: There is close
agreement with classical general relativity away from the
Planck scale. The ultraviolet-infrared tension is avoided
because, although quantum geometry effects are truly enormous
in the Planck regime, \emph{they die extremely quickly.}

\subsection{Generalizations}
\label{s3.2}

In this sub-section I will discuss the fate of singularities and
Planck scale physics in models with increasing generality.

\b \emph{Beyond the big bang and big crunch singularities:} In
section \ref{s3.1} I focused on the big bang and big crunch
singularities where the scale factor goes to zero and the matter
density and curvature diverges at a finite proper time. But for
general matter, new types of singularities can arise even with k=0
and $\Lambda=0$. For example, in the `big-rip' singularity, the
scale factor \emph{diverges} in a finite proper time along with
energy density and pressure. Another possibility is the `sudden
death' singularity where the curvature and/or its derivatives
diverge at a \emph{finite} values of the scale factor (and proper
time). In the spatially homogeneous, isotropic context, if matter
has equation of state $p=p(\rho)$, there is a good control on the
type of singularities that can arise \cite{not}. From a physical
perspective, we can divide them into two broad categories: strong
singularities (for example, the big bang and the big crunch) where
observers would be crushed and tame, weak ones (such as those
resulting from shell-crossing) through which they would be able to
propagate. All possible occurrences have been recently analyzed
using effective equations of LQC \cite{ps}. The result is that all
strong singularities are resolved but not necessarily the weak ones.
Since physics does not stop at the weak singularities, they are
widely regarded as `harmless'. An interesting situation arises in
the context of a phantom model in which general relativity predicts
the occurrence of a big-rip singularity. If one uses the effective
equations of LQC, the energy density remains bounded but the
pressure and the rate of change of the Hubble parameter diverge
\cite{ccvw}. At first this might seem as a problem for LQC. However,
a closer examination \cite{ps} shows that this is a weak singularity
beyond which geodesics can be extended. Thus, in this case the
quantum geometry effects tame a strong singularity and render it
harmless.

\begin{figure}[]
  \begin{center}
    $a)$\hspace{8cm}$b)$
    \includegraphics[width=3.2in,angle=0]{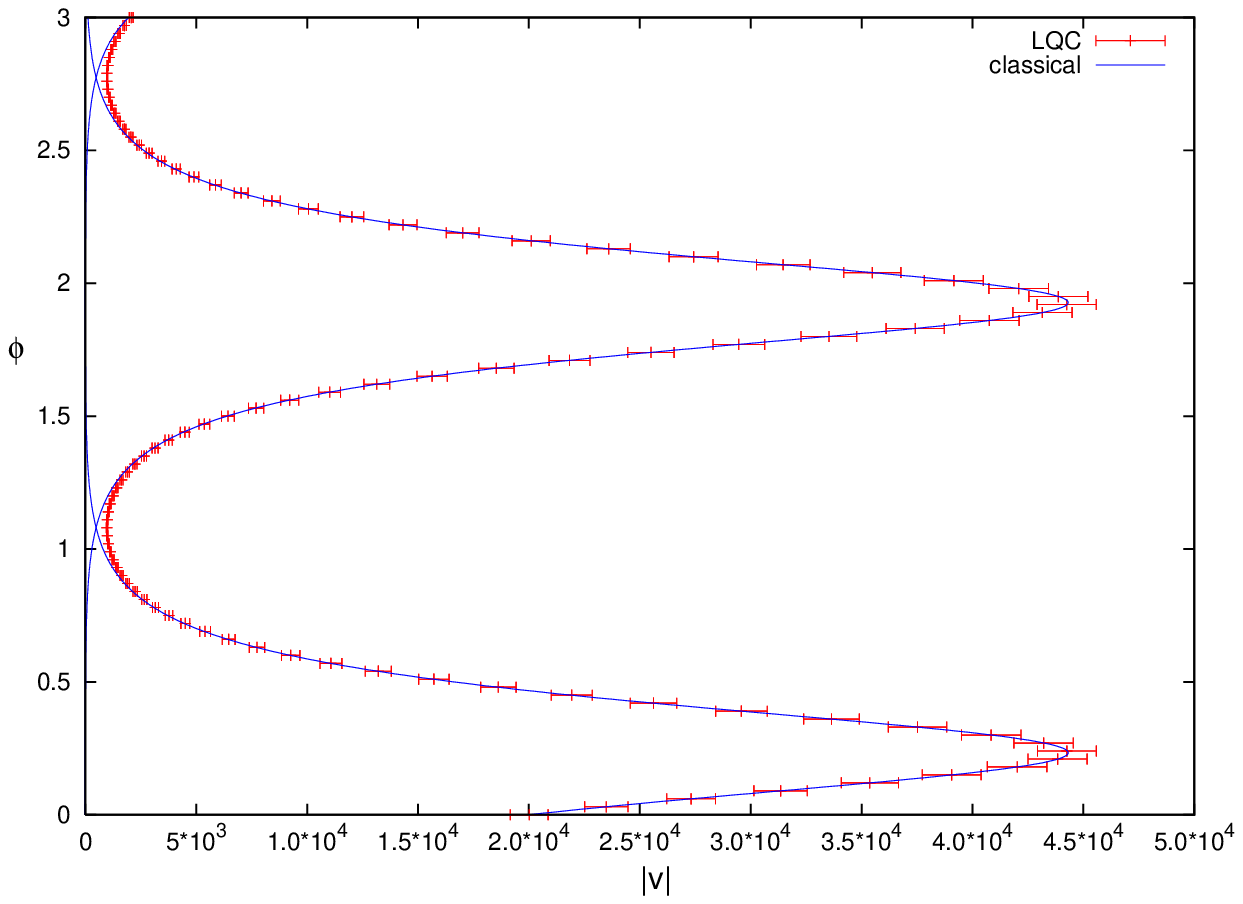}
    \includegraphics[width=3.2in,angle=0]{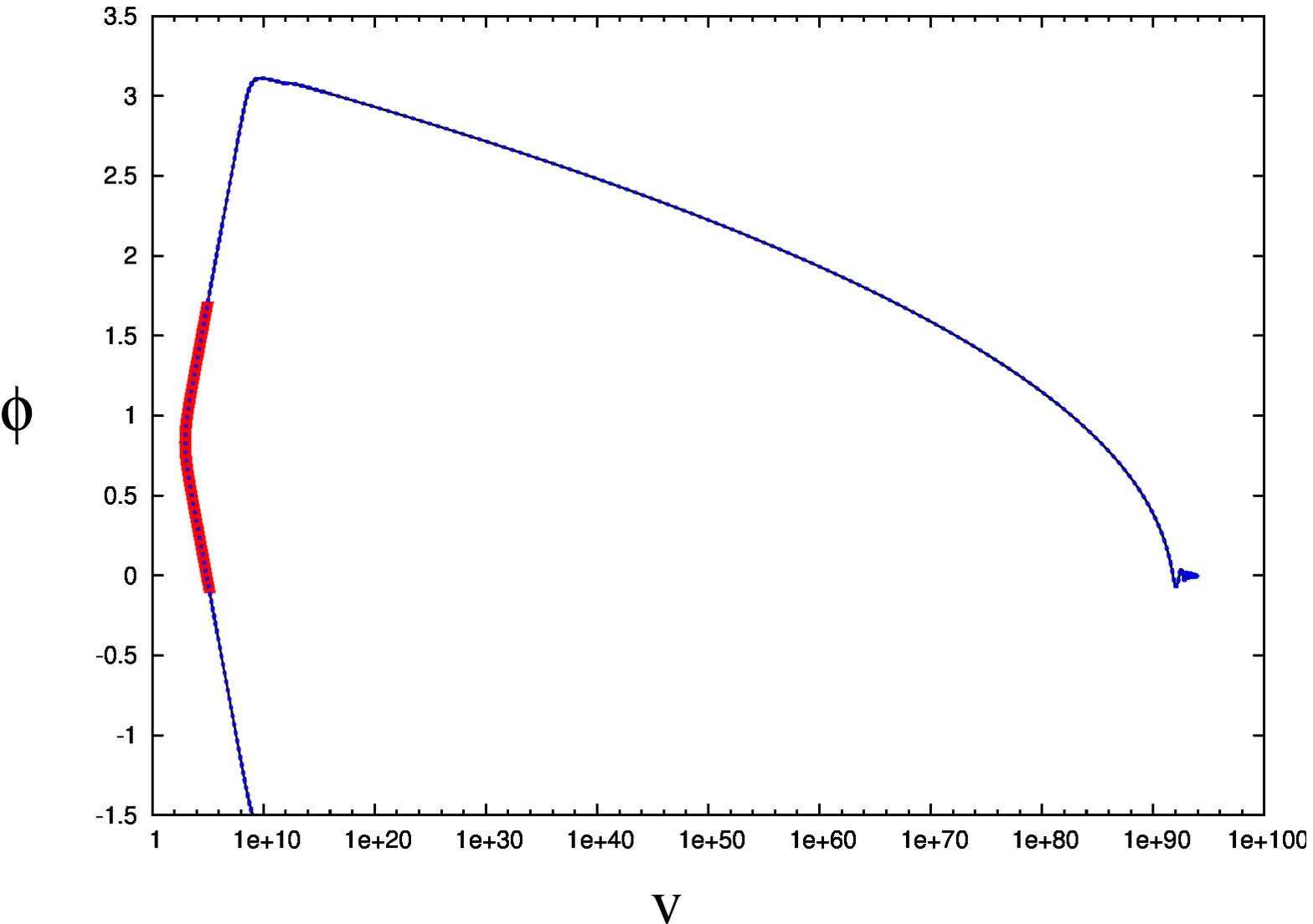}
\caption{ Again, expectation values of the volume operator are
plotted on the x-axis and emergent time $\phi$ on the y-axis.
$a)$ Negative cosmological constant. In LQC, the big
bang and big crunch singularities are resolved and the universe
also undergoes a classical re-collapse. Therefore, the qualitative
behavior is similar to that in k=1, $\Lambda=0$ model of Fig. 2.b.
\,\, $b)$ Inflation with $m^2\phi^2$ potential. LQC reproduces the
general relativistic inflation captured by the long, horizontal part
of the curve that slopes down. However, if we evolve
backward in time using LQC, one obtains the near vertical segment
on the left showing that the big-bang is replaced by a quantum bounce.
Near the bounce, the potential plays a negligible role in the LQC
dynamics.}
\label{fig:lqc2}
\end{center}
\end{figure}

\b \emph{Negative cosmological constant:} The k=0, $\Lambda<0$ model
is discussed in \cite{bp}. In this case, in classical general
relativity the situation is analogous to that in the k=1 model,
discussed above. The universe starts out with a big bang, expands to
a maximum value of the scale factor, at which the positive energy
density of matter is exactly balanced by the negative energy density
in the cosmological constant. If the universe were to expand
further, the total energy density would become negative. But since
the Friedmann equation (\ref{fe}) of classical general relativity
does not allow the total energy density to become negative, the
universe recollapses. In LQC the model has been analyzed in detail
both by solving the quantum Einstein's equations numerically and by
analyzing the solutions of effective equations. Again, for states
which are sharply peaked Gaussians at late times, the effective
equations provide an accurate account of full quantum dynamics,
including the Planck regime. The quantum corrected Friedmann
equation again has the form (\ref{lqc-fe}) where $\rho$ now includes
also the energy density in the cosmological constant; the big bang
and the big crunch are replaced by quantum bounces. The qualitative
picture is thus similar to that in the k=1, $\Lambda=0$ models. The
evolution is (approximately) cyclic. The total density is maximum
($\rho =\rcr$) at the quantum bounce, the universe then expands, and
the total density decreases. When it vanishes, there is a classical
re-collapse, the universe starts contracting, and the total density
increases till it reaches the maximum value, $\rcr$ when there is
again a quantum bounce. The value of $\rcr$ is the same as in the
$\Lambda=0$ case. (See Fig 3.a)

\b \emph{Positive cosmological constant:} In classical general
relativity, the situation in the k=0, $\Lambda >0$ case is similar
to that in the k=0, $\Lambda=0$ case, discussed in section
\ref{s3.1}. But the classical trajectory which starts out at the
big-bang now expands to an infinite volume (so that the energy
density $\rho_\phi$ in the scalar field goes to zero) at a
\emph{finite} value $\phi_{\rm max}$ of the emergent time $\phi$ .
This opens up an interesting possibility in LQC. Because the
evolution in the emergent time $\phi$ is unitary in LQC, one can
continue it beyond the point at which the density goes to zero.
States which are semi-classical in the low $\rho_\phi$ regime again
follow effective trajectories which now naturally extend beyond
$\phi_{\rm max}$. Since $\rho_\phi$ remains bounded, it is
convenient to draw these trajectories in the $\rho_\phi$-$\phi$
plane. They agree with the classical trajectories in the low
$\rho_\phi$ regime and the extension is just the analytical
continuation of classical trajectories. It would be very interesting
to work out physical ramifications of this unforeseen feature of LQC
dynamics. Finally, in this case, LQC predicts that the cosmological
constant has a maximum possible value. As one would expect the bound
is given by the Planck scale. Therefore it is not of
phenomenological interest. However, it is conceptually interesting
that quantum geometry effects imply that $\Lambda$ cannot even in
principle be arbitrarily large.

\b \emph{Inflationary scenarios:} The LQC framework has also been
extended to incorporate an inflaton. Quantum Einstein's equations
have been numerically solved for the $m^2\phi^2$ potential
\cite{aps4}. LQC admits wave functions that are sharply peaked at a
standard inflating trajectory. This is just as one would expect
because in the standard scenario inflation occurs well away from the
Planck regime where general relativity is an excellent approximation
to LQC. When such a wave function is evolved \emph{backward} in
time, one again finds that, rather than following the classical
trajectory into the big bang, the wave function bounces (see Fig.
3.b)). Again, effective equations provide an excellent approximation
to the full quantum equations. Numerical simulations as well as the
effective equations show that the potential has negligible effect
near the big bounce. All the principal features of the LQC evolution
reported above are recovered, including the value of $\rcr$.

\b \emph{Anisotropies and gravitational waves:} Inclusion of
anisotropies leads to Bianchi models. In these models, the Weyl
curvature is not zero because, physically, they admit
gravitational waves. Detailed LQC analysis has been carried out
in Bianchi I and II models at the analytical level
\cite{awe2,awe3,cs2} (and the method extends to all class A
Bianchi models; Bianchi IX, for example, is discussed in
\cite{ew}). It establishes that the big bang singularity is
resolved. Effective equations again show that the matter
density $\rho_{\rm matt}$ is bounded above. However, as one
might expect, the bound now is lower than $\rcr$ because there
is also some energy in the gravitational waves that is not
captured in matter density. In the isotropic models, space-time
curvature is completely encoded in the Ricci scalar. In the
anisotropic case, we also have non-trivial Weyl curvature.
Consequently, there is not just one bounce. Rather, any time a
curvature invariant (more precisely, shear) enters the Planck
regime, quantum geometry effects intervene and dilute it,
thereby avoiding the singularity that would have occurred had
one used general relativity. Thus, the Planck regime is much
richer. However, outside this regime, general relativity is
again an excellent approximation to LQC. Finally, as in the
isotropic case, matter can be chosen to satisfy all energy
conditions. The singularity theorems are transcended because
Einstein's equations are modified just in the right fashion by
LQC.

\b \emph{Beyond homogeneity:} In mathematical general
relativity, the so called Gowdy cosmological models have drawn
much attention because they admit \emph{local, inhomogeneous
degrees of freedom} and yet are sufficiently simple to be
tractable. These models have been studied in detail using LQC
\cite{hybrid}. One can construct a quantum theory of Gowdy
models using conventional field theoretical methods which do
not take into account quantum geometry of LQC. One then finds
that the singularity is not resolved. On the other hand, using
LQC to treat just the homogeneous degrees of freedom already
suffices to resolve singularities, even if the inhomogeneous
modes are treated using more conventional techniques rooted in
Fock quantization. The general expectation is that the
ultra-violet behavior would further improve if all modes are
treated using loop quantum gravity. But already this `hybrid'
scheme introduced by the Madrid group has given some confidence
that \emph{the singularity resolution is not tied to the
homogeneous models.}

\b \emph{General space-like singularities:} In general
relativity there is a conjecture due to Belinskii, Lifshitz and
Khalatnikov (BKL)  which says that as one approaches space-like
singularities in general relativity, `spatial derivatives of
basic fields become sub-dominant relative to the time
derivatives' and dynamics at any spatial point is well
approximated by that of homogeneous models \cite{bkl}. Bianchi
I dynamics plays a dominant role and from time to time there
are transitions from one Bianchi I solution to another,
mediated by a Bianchi II solution. By now there is considerable
support for this conjecture both from rigorous mathematical and
numerical investigations \cite{bb}. This, in turn, provides
support for the hope that the lessons on the quantum nature of
singularities we learned from the Bianchi I and Bianchi II
models may be valid much more generally. In particular, it
suggests that the quantum geometry effects of LQG may well
resolve \emph{generic} space-like, strong curvature
singularities of classical general relativity. These are
precisely the singularities of direct interest to cosmology.
There is now a formulation of the BKL conjecture in the
framework that underlies LQG \cite{ahs}. Together with the
quantization procedure introduced in the Bianchi II models
\cite{awe3} ---which, as remarked above, is directly applicable
to a wide class of homogeneous models--- this form of the BKL
conjecture should enable one to obtain concrete results for
general space-like singularities.

\subsection{Super-inflation and inflation}
\label{s3.3}

So far, I have focused on singularity resolution and the new physics
in the Planck regime. It turns out that LQC has unforeseen
implications also to the inflationary scenarios and phenomena that
occur away from the Planck regime. Because of space limitation, I
will outline only one of these and mention a few other avenues that
are being pursued both by cosmology and LQC communities.

\subsubsection{Inflaton and super-inflation}
\label{3.3.1}

In this sub-section, I will list some basic predictions of LQC which
are rather surprising. These new features arise in the Planck regime
but are important for setting the initial conditions for slow roll
inflation at much lower energies. For simplicity and definiteness,
in the detailed discussion I use the k=0, $\Lambda=0$ isotropic
cosmologies and assume that the kinetic energy of the inflaton is
positive.

As we saw, LQC is generally formulated using volume $v \sim a^3$ of
a fiducial cell (rather than the scale factor $a$), and its
conjugate momentum $b$ \cite{acs,cs1}. On solutions to Einstein's
equations, $b= \gamma H$ where, as before, $\gamma \approx 0.24$ is
the Barbero-Immirzi parameter of LQC and $H = \dot{a}/a$ is the
Hubble parameter. However, LQC modifies the Einstein dynamics and on
solutions to the effective equations we have
\be \label{H} H \,=\, \f{1}{2\gamma\lambda}\, \sin 2\lambda b
\,\,\approx\,\, \f{0.94}{\lp}\, \sin 2\lambda b \ee
where $\lambda^2 \approx 5.17 \lp^2$ is the `area-gap', the
smallest non-zero eigenvalue of the area operator. In LQC $b$
ranges over $(0, \pi/\lambda)$ and general relativity is
recovered in the limit $\lambda \rightarrow 0$.

As I already mentioned, quantum geometry effects modify the
geometric, left side of Einstein's equations. In particular,
the LQC Friedmann equation first emerges as
\be \label{lqc-fe2} \f{\sin^2 \lambda b}{\gamma^2\lambda^2}\,\,
=\,\, \f{8\pi G}{3}\, \rho \,\,\equiv\,\, \f{8\pi G}{3}\,
\big(\f{{\dot\phi}^2}{2} + V(\phi) \big)\, .\ee
To compare with the standard Friedmann equation (\ref{fe}), it is
often convenient to rewrite it in the form (\ref{lqc-fe}) using
(\ref{H}). By inspection it is clear from Eqs (\ref{lqc-fe}),
(\ref{H}) and (\ref{lqc-fe2}) that, away from the Planck regime
---i.e., when $\lambda b \ll 1$ or, $\rho \ll \rho_{\rm Pl}$--- we
recover classical general relativity. However, modifications in the
Planck regime are drastic. Recall that in general relativity, the
Hubble parameter $H$ is large throughout the Planck regime and
diverges at the singularity. By contrast, in LQC $H$ \emph{vanishes}
at the bounce (because $\dot{a} =0$ there) and Eq.(\ref{H}) implies
that it has a \emph{finite}, maximum value, $H= 0.94$. Second, Eq.
(\ref{lqc-fe2}) implies that the density $\rho$ is bounded by $\rcr
\approx 0.41\rho_{\rm Pl}$. Third, if the potential $V(\phi)$ is
bounded below, say $V \ge V_o$, then it follows from (\ref{lqc-fe2})
that ${\dot\phi}^2$ is bounded by $2(\rcr - V_o)$. Fourth, if the
potential grows unboundedly for large $|\phi|$, then $|\phi|$ is
also bounded. For example, for $V= m^2\phi^2/2$, we have
$m|\phi|_{\rm max} = 1.41\sqrt{\rcr}$. Finally, the derivative with
respect to proper time of $b$ and the Hubble parameter are now given
by
\be \label{dot} \dot{b} = - 4\pi \gamma G\, {\dot\phi}^2 \qquad
{\rm and}\qquad \dot{H} = -4\pi G\, (\cos 2\lambda b)\,
\dot\phi^2 \ee
As a consequence, $b$ \emph{decreases monotonically in every
solution}, starting from $b = \pi/\lambda$ and ending with
$b=0$ (the bounce occurs at the mid-point, $b = \pi/2\lambda$).
When the potential is bounded below, $|\dot{H}|$ is also
bounded by $10.3/\lp^2$. Thus, a large number of physical
quantities which are unbounded in general relativity cannot
exceed certain finite, maximum values in LQC.

In particular, these results imply that if the potential is bounded
below, the matter density and curvature can not diverge anywhere on
LQC space-times. One can also show that a solution where our fixed
fiducial cell has a finite volume initially cannot evolve to a
configuration where the volume becomes zero. Thus, \emph{the LQC
solutions are everywhere regular irrespective of whether one focuses
on matter density, curvature or the scale factor.} Finally, every
solution undergoes precisely one bounce at $b= \pi/2\lambda$ where
the Hubble parameter vanishes because of Eq. (\ref{H}) and the
density reaches its maximum value $\rcr$ because of Eq.
(\ref{lqc-fe}).

Let me now turn to \emph{super-inflation} (see \cite{mb2} and
especially \cite{ps2}). At the bounce point, we have
$b=\pi/2\lambda$ whence (\ref{H}) implies that $H=0$ and it follows
from (\ref{dot}) that $\dot{H}$ is positive. The universe expands to
the future of the bounce point and contracts in its past. Eq
(\ref{dot}) shows that $\dot{H}$ continues to be positive till $b$
has decreased to $b= \pi/4\lambda$ at which point it vanishes (and
then it becomes and remains negative as $v$ increases monotonically
to infinity). Thus, \emph{every LQC solution has a
super-inflationary phase from} $b= \pi/2\lambda$ to
$b=\pi/4\lambda$. Eqs (\ref{H}) and (\ref{dot}) imply that at the
beginning of this phase, the Hubble parameter $H$ vanishes and
$\dot{H}$ is very large, while at the end of this phase $H$ assumes
its maximum value and $\dot{H}$ vanishes. Note that the occurrence
of this phase is universal; it exists even if the inflaton potential
is zero! \emph{This is a robust feature of LQC that has no analog in
general relativity.} It is natural to ask if this quantum geometry
driven super-inflation could be a substitute for the standard
inflation. If so, we would not have to invoke \emph{any} potential!
Unfortunately the answer is in the negative. Detailed considerations
show that in absence of a potential, this phase is very short lived
in proper time and therefore does not yield enough e-foldings to be
a viable substitute for the standard inflation.

However, it was recently realized that in presence of the
`standard' potentials used in inflation, this phase has an
unforeseen consequence: it naturally funnels dynamical
trajectories to initial conditions that \emph{lead to a long
slow roll inflation with at least 68 e-foldings} \cite{as}.

\subsubsection{LQC and inflation}
\label{3.3.2}

Inflationary scenarios have had an extraordinary success especially
in accounting for structure formation. This success brings added
urgency to a long standing question: Does a sufficiently long, slow
roll inflation require fine tuning of initial conditions or does it
occur generically in a given theoretical paradigm? (See e.g.
\cite{hw,klm,hhh,gt}). Such a slow roll requires that initially the
inflaton must be correspondingly high-up in the potential. How did
it get there? Is it essential to invoke some rare quantum
fluctuations to account for the required initial conditions because
the a priori probability for their occurrence is low? Or, is a
sufficiently long, slow roll inflation robust in the sense that it
is realized in `almost all' dynamical trajectories of the given
theory?

To make these questions precise, one needs a stream-lined framework
to calculate probabilities of various occurrences \emph{within a
given theory}. A mathematically natural framework to carry out this
analysis was introduced over two decades ago (see, e.g.,
\cite{ghs,dp,hp}). It invokes Laplace's principle of indifference
\cite{psdl} to calculate the \emph{a priori} probabilities for
various occurrences. More precisely, the idea is to use (a flat
probability distribution $P(s)=1$ and) the canonical Liouville
measure $\dd \mu_{\L}$ on the space ${\S}$ of solutions $s$ of the
theory under consideration to calculate the \emph{relative volumes}
in ${\S}$ occupied by solutions with desired properties \cite{ghs}.
In our case, then, the a priori probability is given by the
\emph{fractional} Liouville volume occupied by the sub-space of
solutions in which a sufficiently long, slow roll inflation occurs.
Further physical input can provide a sharper probability
distribution $P(s)$ and a more reliable likelihood than the `bare' a
priori probability. However, a priori probabilities can be directly
useful if they are very low or very high. In these cases, it would
be an especially heavy burden on the fundamental theory to come up
with the physical input that significantly alters them.

The task, then, is to calculate the \emph{a priori} probability that
there is inflation with at least 68 e-foldings. As in general
relativity, this can be done by introducing the natural Liouville
measure on the space ${\S}$ of solutions and calculating the
fractional Liouville volume of ${\S}$ occupied by solutions with
adequate number of e-foldings. In general relativity, it has been
argued \cite{gt} that the resulting a priori probability for
obtaining 68 or more e-foldings is suppressed by a factor larger
than $e^{-204}$. What is the situation in LQC? Now the equations of
motion are
\be \label{ddot}\ddot{v} = \f{24\pi G}{\rcr}\, v\,\big[(\rho -V)^2 +
V(\rcr -V)\big]\quad\quad {\rm and} \quad\quad \ddot{\phi} +3H
\dot{\phi} + V_{,\phi} =0 \ee
where $3H = 3\dot{a}/a = \dot{v}/v$, and $v, \phi$ are, in addition,
subject to the LQC-modified Friedmann equation (\ref{lqc-fe}). It is
simplest to write the natural Liouville measure on the space ${\S}$
of solutions in terms of values $\phi_{\B},v_{\B},$ of the scalar
field and the volume at the bounce:
\be \dd\hat\mu_{\L} =   \f{\sqrt{3\pi}}{\lambda}\,\, \big[1-
F_{\B}\big]^{\f{1}{2}}\,\, {\dd} \phi_{\B}\, {\dd} {v}_{\B} \ee
where $F_{\B} = V(\phi_{\B})/\rcr$ is the fraction of the total
density that is in the potential energy at the bounce. Furthermore,
in LQC $|\phi|$ takes values in the \emph{bounded} interval $(0,\,
\phi_{\rm max}=0.9\sqrt{\rho_{\rm Pl}}/m)$. The question now is:
With respect to this measure, what is the fractional volume in
${\S}$ occupied by solutions with at least 68 e-folds of slow roll
inflation?

Recall that the Hubble parameter $H$ takes its maximum value at the
end of super-inflation and the friction term in the equation of
motion for $\phi$ is given by $H/m^2$. The phenomenologically
preferred value of $m$ is of the order $10^{-6}\,M_{\rm Pl}$.
Therefore, at the end of super-inflation, the friction term is
necessarily large and initial conditions are naturally set for a
long slow roll. To calculate the probabilities, let us use the
phenomenological value of $m$. One can perform detailed analytical
calculations using usual approximations and check their validity via
high precision numerical simulations. One finds that 68 or more
e-foldings are guaranteed if the fraction $F(\B)$ is larger than
$1.4\times 10^{-11}$ \cite{as}. This in turn implies that the a
priori probability of a slow roll with at least 68 e-foldings is
greater than 0.99. This conclusion is quite robust: One can change
the inflaton mass by a couple of orders of magnitude or add a
quartic term to the potential (with phenomenological bounds on the
coupling constant). The probability remains greater than 0.99.

Thus, even when one uses the same methods that were used for general
relativity in \cite{gt}, the conclusion in LQC is \emph{opposite} of
that in \cite{gt}. This comes about because, unlike in general
relativity, $\phi, \dot\phi$ are bounded in LQC; all solutions are
singularity free and undergo a quantum bounce; and the bounce is
followed by a period of super-inflation which funnels dynamical
trajectories to initial conditions that are well suited
for a long slow roll.\\

There are several other phenomenological implications of super
inflation, and more generally, of the LQC quantum corrections to
Einstein's equations, particularly on scalar and tensor
perturbations (see e.g. \cite{nott,gre1,gre2}). I discussed some of
these in my talk. Unfortunately, space limitation does not allow me
to discuss these developments here.

\section{Outlook}
\label{s4}

Singularities of general relativity are perhaps the most promising
gates to physics beyond Einstein. They provide a fertile conceptual
and technical ground in our search of a new paradigm in cosmology as
well as fundamental physics. Consider some of the deepest conceptual
questions we face today: the issue of the Beginning and the end End,
the arrow of time, and the puzzle of black hole information loss.
Their resolutions hinge on the true nature of space-time
singularities. In my view, considerable amount of contemporary
confusion about such questions arises from our explicit or implicit
insistence that singularities of general relativity are true
boundaries of space-time; that we can trust causal structure all the
way to these singularities; that notions such as event horizons are
absolute even though changes in the metric in a Planck scale
neighborhood of the singularity can move event horizons dramatically
or even make them disappear altogether \cite{ph}.

LQG is well suited to address the issue of the fate of classical
singularities in quantum gravity because it is fully
non-perturbative and does not pre-suppose that we have a smooth
geometry all the way up to the big-bang and big crunch
\cite{alrev,crbook,ttbook}. Therefore LQC has been used to address
many long standing cosmological questions in detail
\cite{mb-rev,aa-badhonef,cs1}. The scalar field serves as emergent
time. Strong curvature singularities of classical general relativity
are either resolved \cite{aps3,apsv,kv,bp,ap,aps4,acs,cs1} or
converted to harmless weak singularities by the quantum geometry
effects \cite{ps}. In particular, the big bang and the big crunch
are naturally replaced by quantum bounces. On the `other side' of
the bounce there is again a large universe. General relativity is an
excellent approximation to quantum dynamics once the matter density
falls below one percent of the Planck density. Thus, LQC
successfully meets both the `ultra-violet' and `infra-red'
challenges. Furthermore results obtained in a number of models using
distinct methods re-enforce one another. One is therefore led to
take at least the qualitative findings seriously: \emph{Big bang is
not the Beginning nor the big crunch the End.} Quantum space-times
could be vastly larger than what general relativity had us believe!

How can the quantum space-times of LQC manage to be significantly
larger than those in general relativity when those in the \WDW
theory are not? Main departures from the \WDW theory occur due to
\emph{quantum geometry effects} of LQG. There is no fine tuning of
initial conditions, nor a boundary condition at the singularity,
postulated from outside. Furthermore, matter can satisfy all the
standard energy conditions. Why then does the LQC singularity
resolution not contradict the standard singularity theorems of
Penrose, Hawking and others? These theorems are inapplicable because
\emph{the left hand side} of the classical Einstein's equations is
modified by the quantum geometry corrections of LQC. What about the
more recent singularity theorems that Borde, Guth and Vilenkin
\cite{bgv} proved in the context of inflation? They are not tied to
Einstein's equations. But, motivated by the eternal inflationary
scenario, they assume that the expansion remains positive if we
recede in the past along any geodesic. Because of the pre-big-bang
contracting phase, this assumption is violated in the LQC effective
theory.

A qualitative picture that emerges is that the non-perturbative
quantum geometry corrections create a \emph{`repulsive'} force.
While this force is negligible under normal conditions, it dominates
when curvature approaches the Planck scale and can halt the collapse
that would classically have led to a singularity. In this respect,
there is a curious similarity with the situation in the stellar
collapse where a new repulsive force comes into play when the core
approaches a critical density, halting further collapse and leading
to stable white dwarfs and neutron stars. This force, with its
origin in the Fermi-Dirac statistics, is \emph{associated with the
quantum nature of matter}. However, if the total mass of the star is
larger than, say, $5$ solar masses, classical gravity overwhelms
this force. The new repulsive force of LQC is \emph{associated with
the quantum nature of geometry}. It comes into play only near Planck
densities but is then so strong that it can counter the classical,
gravitational attraction, irrespective of how large the collapsing
mass is. It is this force that prevents the formation of
singularities.

At first one might think that, since quantum gravity effects concern
only a tiny region, whatever they may be, their influence on the
global properties of space-time should be negligible whence they
would have almost no bearing on the issue of the Beginning and the
End. However, as we saw, once the singularity is resolved, vast new
regions appear on the `other side.' New possibilities open up that
were totally unforeseen in general relativity \cite{as}. First,
matter density, curvature scalars, and the Hubble parameter
\emph{are all bounded}. Second, even in absence of a potential,
there is a robust super-inflationary phase immediately after the
quantum bounce \cite{mb2,ps2}. As we discussed in section \ref{s3.3}
this novel phase has interesting consequences. In particular, it
naturally leads to the phenomenon of \emph{Hubble funneling}: on
every solution the Hubble parameter is driven to its largest
possible value at the end of super-inflation. As a consequence, if
there is a potential such as $m^2\phi^2$, initial conditions are
naturally set for a long slow roll inflation \cite{as}. One does not
have to appeal to rare quantum fluctuations to explain how the
inflaton managed to get sufficiently high in the potential to seed a
slow roll inflation leading to 68 or more e-foldings. Finally,
\emph{these conclusions are insensitive to what happened in the
pre-big-bang branch.}

Another major direction of current work concerns perturbations that
seed structure formation. In the new paradigm provided by LQC, one
is led to re-examine the issue of initial conditions for these
perturbations. We no longer have to specify them at a singularity.
It is most natural to set them in the infinite past where matter
density and curvatures tend to zero, whence it makes direct physical
sense to speak of vacuum fluctuations. The LQC paradigm leads us to
re-asses the old questions from entirely new angles (see, e.g.,
\cite{nott,gre1,gre2}). This opens up a wealth of interesting
challenges and LQC provides novel ideas and tools to meet them.
\bigskip

\textbf{Acknowledgments:} I would like to thank Alex Corichi, Tomasz
Pawlowski, Param Singh, David Sloan, Victor Taveras, Kevin
Vandersloot and Edward Wilson-Ewing for collaboration and
discussions. This work was supported in part by the NSF grant
PHY0854743, the Alexander von Humboldt Foundation, The George A. and
Margaret M. Downsbrough Endowment and the Eberly research funds of
Penn State.

\end{document}